\PassOptionsToPackage{prologue,dvipsnames}{xcolor}
\documentclass[sigconf]{acmart}

\AtBeginDocument{%
  \providecommand\BibTeX{{%
    \normalfont B\kern-0.5em{\scshape i\kern-0.25em b}\kern-0.8em\TeX}}}

\setcopyright{acmcopyright}
\copyrightyear{2023}
\acmYear{2023}
\acmDOI{XXXXXXX.XXXXXXX}

\acmPrice{15.00}
\acmISBN{978-1-4503-XXXX-X/18/06}

\usepackage{multirow}
\usepackage{color}
\usepackage{colortbl}
\usepackage{xcolor}
\renewcommand\footnotetextcopyrightpermission[1]{}
\begin{document}

\title{A Dual Stealthy Backdoor: From Both Spatial and Frequency Perspectives}


\author{Yudong Gao}
\authornotemark[1]
\email{YudongGao0504@163.com}
\affiliation{%
  \institution{China University of Petroleum (East China)}
  \country{China}
}

\author{Honglong Chen}
\authornotemark[2]
\affiliation{%
  \institution{China University of Petroleum (East China)}
  \country{China}
}
\author{Peng Sun}
\authornotemark[3]
\affiliation{%
  \institution{Hunan University}
  \country{China}
}

\author{Junjian Li}
\authornotemark[4]
\affiliation{%
  \institution{China University of Petroleum (East China)}
  \country{China}
}

\author{AnQing Zhang}
\authornotemark[5]
\affiliation{%
  \institution{China University of Petroleum (East China)}
  \country{China}
}

\author{Zhibo Wang}
\authornotemark[6]
\affiliation{%
  \institution{Zhejiang University}
  \country{China}
}
\renewcommand{\shortauthors}{none}
\settopmatter{printacmref=false} 
\begin{abstract}
 Backdoor attacks pose serious security threats to deep neural networks (DNNs). Backdoored models make arbitrarily (targeted) incorrect predictions on inputs embedded with well-designed triggers while behaving normally on clean inputs. Many works have explored the invisibility of backdoor triggers to improve attack stealthiness. However, most of them only consider the invisibility in the spatial domain without explicitly accounting for the generation of invisible triggers in the frequency domain, making the generated poisoned images be easily detected by recent defense methods. To address this issue, in this paper, we propose a \textbf{DU}al stealthy \textbf{BA}ckdoor attack method named \textbf{DUBA}, which simultaneously considers the invisibility of triggers in both the spatial and frequency domains, to achieve desirable attack performance, while ensuring strong stealthiness. Specifically, we first use Discrete Wavelet Transform to embed the high-frequency information of the trigger image into the clean image to ensure attack effectiveness. Then, to attain strong stealthiness, we incorporate Fourier Transform and Discrete Cosine Transform to mix the poisoned image and clean image in the frequency domain. Moreover, the proposed DUBA adopts a novel attack strategy, in which the model is trained with weak triggers and attacked with strong triggers to further enhance the attack performance and stealthiness. We extensively evaluate DUBA against popular image classifiers on four datasets. The results demonstrate that it significantly outperforms the state-of-the-art backdoor attacks in terms of the attack success rate and stealthiness.
\end{abstract}





\maketitle

\section{Introduction}
Deep neural networks (DNNs) have made great achievements in many fields, such as image classification~\cite{he2016deep}, image segmentation~\cite{feng2022fiba}, and target recognition~\cite{peng2021conformer}. Despite the remarkable success, DNNs face various security threats since the models are usually trained on datasets labeled by third-parties or even outsourced for training. Recent studies have shown that DNNs are vulnerable to backdoor attacks~\cite{gu2017badnets}, where an adversary intentionally manipulates a part of the training data or modifies the model parameters to make the model behave correctly on clean data but make arbitrarily (targeted) incorrect predictions on poisoned data. Backdoor attacks pose serious security threats to deep learning systems, especially in security-sensitive applications (e.g., autonomous driving~\cite{nguyen2021wanet}).

\begin{figure}[h]
	\centering
	\includegraphics[height=4.29cm]{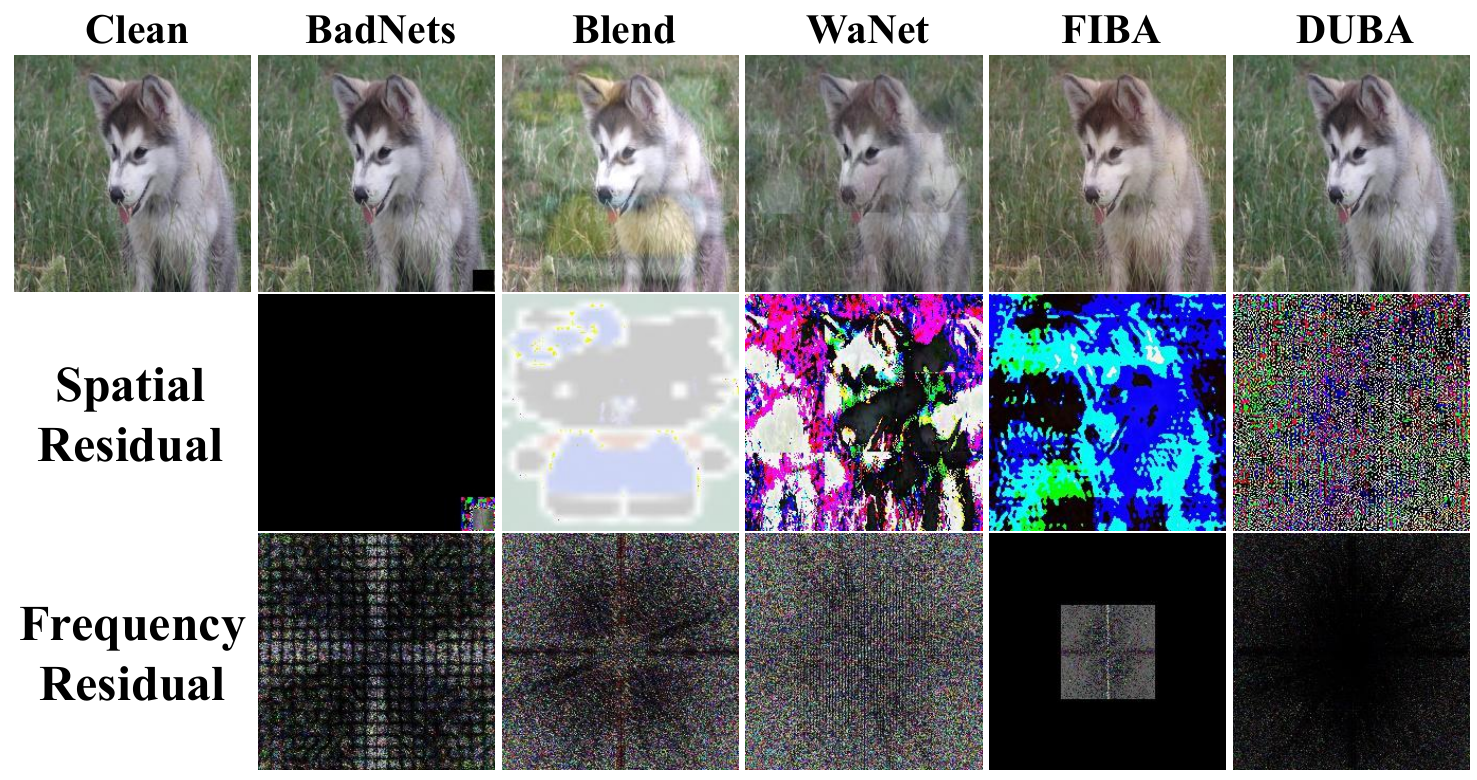}
    \caption{Visualization of images poisoned by different attacks. The images in the first row are: the clean image, images poisoned by BadNets~\protect\cite{gu2017badnets}, Blend~\protect\cite{chen2017targeted}, WaNet~\protect\cite{nguyen2021wanet}, FIBA~\protect\cite{feng2022fiba}, and the proposed DUBA. The second row is the corresponding residual images in the spatial domain and the bottom is the residual images in the frequency domain.}
	\label{visuall}
\end{figure} 
Most existing backdoor attacks craft backdoor triggers in the spatial domain~\cite{gu2017badnets,chen2017targeted,li2021invisible}. However, as shown in Figure \ref{visuall}, with the visible triggers, early backdoor attacks (e.g., BadNets~\cite{gu2017badnets} and Blend~\cite{chen2017targeted}) can be easily detected and removed. To enhance attack stealthiness, recent works consider generating invisible triggers through image steganography~\cite{li2021invisible}, disorting fields~\cite{nguyen2021wanet} and so on. Nevertheless, these methods only consider the invisibility of triggers in the spatial domain but not the frequency domain. Consequently, the generated backdoored images can be easily identified by typical image classification models that employ the Fourier transform as a part of the task pipeline~\cite{zhang2021deep,yan2019fourier}. More importantly, researchers have proposed effective backdoor defenses from frequency perspective. For example, FTD~\cite{zeng2021rethinking} demonstrated that most backdoor triggers are high-frequency semantics, and it trained a DNN model that classifies images in the frequency domain to effectively defend against most attacks. That is, most backdoors are perceptible in the frequency domain. Therefore, it is essential to ensure the trigger's invisibility in the both the spatial and frequency domains to launch a powerful yet stealthy backdooor attack. 

A few recent works began to study how to implant stealthy backdoor attacks from frequency perspective. For example, the work in ~\cite{zeng2021rethinking} employs a low-pass filter to implant a backdoor invisible in the frequency domain while visible in the spatial domain. 

Motivated by the above discussions, in this paper, we propose a \textbf{DU}al stealthy \textbf{BA}ckdoor attack called \textbf{DUBA}, which crafts invisible triggers in both the spatial and frequency domains while achieving desirable attack performance. Specifically, we first embed the high-frequency information of the trigger image into the clean image by discrete wavelet transform (DWT), yielding the initial poisoned image. Then, to ensure strong stealthiness, we fuse the initial poisoned image (which has high-frequency triggers) with the clean image in the Fourier and Cosine transform domains. Furthermore, we propose an attack strategy, which can greatly reduce the embedded high-frequency information of the trigger images and randomly mask more parts of the trigger images in the training phase to make the victim model better learn the triggers.

The major contributions of this paper are summarized as follows:
\begin{itemize}
	\item We design a \textbf{DU}al stealthy \textbf{BA}ckdoor attack named DUBA that achieves desirable invisibility in both spatial and frequency domains by embedding high-frequency trigger information through DWT and smoothing it in Fourier Transform and Cosine Transform domains.   
	\item We propose a novel attack strategy for DUBA where the model is trained with weak triggers and attacked with strong triggers to attain satisfactory attack performance while ensuring stealthiness.
	\item We conduct extensive experimental evaluation of DUBA on four datasets and popular models. The results demonstrate its outstanding performance in terms of both attack effectiveness and stealthiness. 
\end{itemize}

\section{Related Work}
\subsection{Backdoor Attack}
Backdoor attack has drawn wide attention since its introduction~\cite{saha2020hidden}. According to the trigger generation method, existing backdoor attacks can be roughly divided into two categories, i.e., spatial domain backdoors and frequency domain backdoors.

\noindent\textbf{Spatial Domain Backdoors.} \quad Researchers in BadNets~\cite{gu2017badnets} first reveals the existence of backdoors in DNNs. This attack embeds a visible square in the bottom right corner of the clean image and manipulates the associated label to the target label. Then, the backdoor can be injected to the model after training on the poisoned data. In the inference phase, input images with the same trigger will be misclassified into the attacker-chosen target label. Inspired by BadNets, researchers have also investigated other backdoor attacks. Blend~\cite{chen2017targeted} advocates image blending backdoors, whereas another work~\cite{fix} employs a fixed watermark as a trigger to insert backdoors. However, these early backdoors are all visually visible and thus can be easily detected and removed. Therefore, how to generate and implant visually invisible backdoors has recently become a hot research topic. For example, ISSBA~\cite{li2021invisible} embeds trigger information by steganography; WaNet~\cite{nguyen2021wanet} crafts triggers by distorting fields; and LIRA~\cite{doan2021lira} searches for triggers in a highly nonlinear parameter space. Though these methods can successfully generate invisible triggers and bypass mainstream backdoor defenses, none of them explicitly account for the characteristics of the image in the frequency domain. Thus, these backdoor attacks can be easily detected by models empowered by Fourier transform (which is often employed as a part of the task pipeline) or the frequency-oriented defense methods.

\noindent\textbf{Frequency Domain Backdoors.} \quad Recently, ~\cite{zeng2021rethinking} starts to explore backdoor attacks in the frequency domain. To avoid high-frequency artifacts after the Discrete Cosine Transform (DCT)~\cite{cintra2011dct}, it applies a low-pass filter to generate a smooth trigger. However, this method yields visible artifacts in the spatial domain. FIBA~\cite{feng2022fiba} crafts triggers in the frequency domain by mixing the low-frequency components of two images after Fast Fourier Transform (FFT)~\cite{moreland2003fft}, which is visually imperceptible in spatial domain but still visible in frequency domain. Another work FTROJAN~\cite{wang2022invisible} first transforms the clean image with YUV or UV (two color coding methods), then applies DCT with modifications on the high-frequency or mid-frequency components to generate poisoned image. However, the transformations required and the frequency components to be modified differ across different images thus increasing the computation overheads. Moreover, the trigger generated in FTROJAN is also visible in the frequency domain. 
\subsection{Backdoor Defense}
To defend DNNs against various backdoor attacks, researchers have proposed many defense methods accordingly~\cite{li2021neural,xu2020defending}. Generally, backdoor defenses can be categorized into input-based, model-based, and output-based methods.

\noindent\textbf{Input-based Defenses.} \quad Input-based defenses focus on input abnormalities~\cite{chou2020sentinet,zeng2021rethinking}. Grad-Cam~\cite{selvaraju2017grad} uses a saliency map to dissect the regions of the input image that the model focuses on. If the model does not focus on the object or keeps focusing on the same region, the image is considered as poisoned. FTD~\cite{zeng2021rethinking} employs DCT to distinguish whether the input image has high-frequency artifacts. They design a DNN-based discriminator to classify images with high-frequency artifacts as poisoned images. 

\noindent\textbf{Model-based Defenses.}\quad They focus on the investigation of the victim model~\cite{kolouri2020universal}. Fine-Pruning~\cite{liu2018fine} mitigates backdoors by pruning dormant neurons since it is likely that these neurons provide specialized support to backdoors. Neural Cleanse~\cite{wang2019neural} identifies whether there is a backdoor in the model by reverse engineering the triggers and utilizes anomaly detection to determine the most potential backdoor.

\noindent\textbf{Output-based Defenses.}\quad Defense methods of this type often observe output anomalies~\cite{huang2020one,gao2019strip}. STRIP~\cite{gao2019strip} superimposes various image patterns on the suspicious image to observe its output. Higher poisoning odds yield lower output randomness. To circumvent the existing backdoor defenses from both spatial and frequency domains, in this work, we aim to craft a powerful backdoor attack that is invisible in both domains. 

\begin{figure*}[]
	\centering
	\includegraphics[height=5.5cm,width=16.5cm]{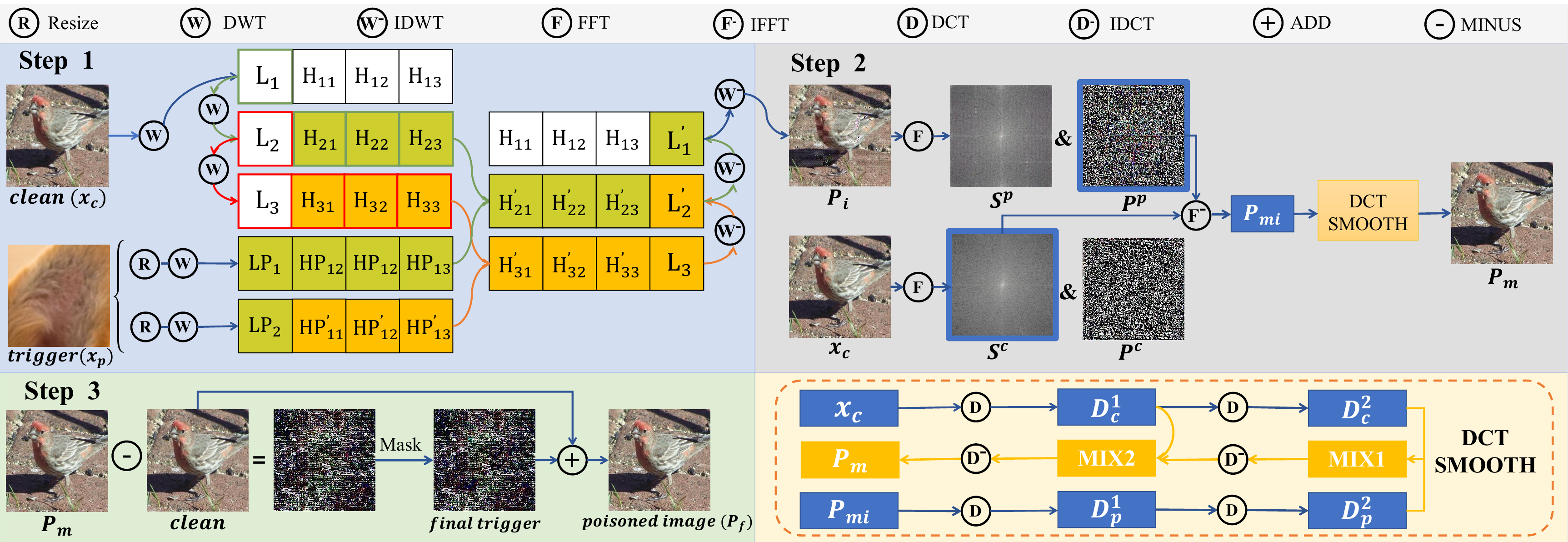}
	\caption{Schematic diagram of DUBA. Step 1: embedding the high-frequency information of the trigger image into the clean image by DWT. Note that the initial trigger $x_p$ is a randomly selected image. Step 2: smoothing the high-frequency trigger in the FFT and DCT domains. Step 3: random trigger masking. In the training phase, we embed weaker high-frequency triggers and mask more pixels of the trigger image. In the attack phase, we use stronger high-frequency triggers to launch the attack. }
 \label{schematic}
\end{figure*} 

\section{Methodology}
\subsection{Threat Model}
\noindent\textbf{Attacker’s Capabilities.} \quad In the training phase, following prior studies~\cite{li2021invisible,DBLP:conf/ijcai/ZhongQZ22}, we consider that the attacker is only allowed to tamper with a part of the training data, but has no access to other model training components (e.g., the victim model architecture and loss function). In the inference phase, the attacker can only manipulate the input images (i.e., embedding the crafted backdoor trigger). Such a threat model can be seen in many real-world scenarios like the outsourcing of model training to third-parties.

\noindent\textbf{Attacker’s Goals.}\quad Generally, an effective backdoor attack should mislead the models into making arbitrarily (targeted) incorrect predictions on the tampered testing images without compromising the models' performance on normal inputs. Furthermore, a powerful backdoor attack should satisfy the following two objectives:
\begin{itemize}
	\item Invisibility. The poisoned images should be invisible in both spatial and frequency domains. 
	\item Robustness. The backdoor attack can successfully circumvent state-of-the-art defense methods. 
\end{itemize}
\subsection{Problem Formulation}
We focus on the typical supervised image classification, which is widely used in face recognition, traffic signal recognition and other security-sensitive fields. Formally, the image classification can be described as a mapping function $f_\theta: \mathcal{X} \rightarrow \mathcal{C}$, where  $\mathcal{X}$ is the input domain and $\mathcal{C}$ is the set of target classes. In the image classification, the model parameters $\theta$ can be learned from the training dataset $D_{\text {train }}=\left\{\left({x}_{{i}}, {y}_{{i}}\right)\right\}_{{i}=1}^{{N}}$ of $N$ data samples, where $x_i \in \mathcal{X}$,
$y_i \in \mathcal{C}$. The core of backdoor attacks is to craft a set of posioned data samples $D_{\text {poison }}=\left\{\left({T}\left({x}_{{i}}\right), \gamma\left({y}_{{i}}\right)\right)\right\}_{{i}=1}^{{M}}$, where ${T}(\cdot)$ denotes the trigger implantation method and $\gamma(\cdot) \in \mathcal{C}$ represents the designated target label. Specifically, $\gamma\left(y_i\right)=c$, where c is a constant, stands for the  All-to-One attack, while $\gamma\left(y_i\right)=y_i+1$ represents All-to-All attack. In summary, backdoor attacks aim to manipulate a subset of the training data (note that $M \textless N$ and $\partial  = \frac{M}{N}$ is the poisoning ratio) by injecting adversarial triggers such that the model trained on the tampered dataset yields the following behaviors when deployed:
\begin{equation}
	{f_\theta }\left( {{x_i}} \right) = {y_i} ,\quad {f_\theta }\left( {T\left( {{x_i}} \right)} \right) = \gamma \left( {{y_i}} \right).
\end{equation}

In this paper, we focus on designing the  trigger implantation method $T(\cdot)$. 

\subsection{The Proposed Attack}
\textbf{Overview of DUBA.} \quad Figure \ref{schematic} shows the framework of DUBA, which is composed of three steps and an attack strategy. First, to attain desirable backdoor attack performance, we employ DWT to embed the high-frequency information of a fixed trigger image into the clean image to generate the initial poisoned image. Second, to ensure strong stealthiness in both spatial and frequency domains, we incorporate FFT and DCT to mix the initial poisoned image with clean image to generate the intermediate poisoned image. Third, to ensure that the victim model learns the backdoor scattered over the entire poisoned image while ensuring good invisibility, we propose to randomly mask the trigger of intermediate poisoned image to get the final poisoned image. Besides, we propose an attack strategy where the victim model is trained with weak triggers and attacked with strong triggers to achieve higher attack success rates (ASRs).

\noindent\textbf{Step 1: High-Frequency Information Embedding.} \quad Inspired by prior works~\cite{zeng2021rethinking,nguyen2021wanet} that utilize high-frequency semantic information for trigger embedding, we propose to extract the high-frequency information from a fixed image (different from the image to be poisoned) as the initial trigger. Specifically, we employ DWT for high-frequency information extraction given that DWT can finely dissect images at high frequencies but coarsely analyze images at low frequencies. Formally, as shown in Step 1 in Figure \ref{schematic}, given a clean image ${x_c}$ and a random initial trigger image ${x_p}$, the idea is to embed the high-frequency part of ${x_p}$ into the deep high-frequency region of ${x_c}$. DWT decomposes image ${x}$ into one low-frequency part and three high-frequency parts, which are represented as:
\begin{equation}
	W\left( {{x}} \right) = \left\{ {{L},{H_{1}},{H_{2}},{H_{3}}} \right\},
\end{equation}
where $L$ represents the low-frequency approximate component, while $H_{1}$, $H_{2}$, and $H_{3}$ denote the high-frequency components at the vertical, diagonal, and horizontal directions, respectively. Accordingly, the image can also be recovered by inverse discrete wavelet transform (IDWT) ${W^ - }$:
\begin{equation}
{W^ - }\left( {L,{H_1},{H_2},{H_3}} \right) = x.
\end{equation}

To embed a sufficiently hidden trigger, we apply three DWTs to clean image ${x_c}$, which is expressed as:
\begin{equation}
	\left\{ {{L_{i + 1}},{H_{i + 1,1}},{H_{i + 1,2}},{H_{i + 1,3}}} \right\} = W\left( {{L_i}} \right),i=0,1,2,
\end{equation}
where $L_0$ equals $x_c$ and $i$ stands for the $i$-th DWT. Then, we apply one DWT to trigger image ${x_p}$. Note that since the image size will change after DWT and we perform three DWTs on ${x_c}$, trigger image $x_p$ first needs to be resized into two different sizes, which are referred to as ${x_{p1}}$ and ${x_{p2}}$. Then, for both ${x_{p1}}$ and ${x_{p2}}$, we apply DWT once to obtain two different high-frequency trigger information.
\begin{equation}
\begin{array}{l}
	\left\{ {L{P_1},H{P_{1,1}},H{P_{1,2}},H{P_{1,3}}} \right\} = W({x_{p1}})\\
	\left\{ {LP_1^{'},HP_{1,1}^{'},HP_{1,2}^{'},HP_{1,3}^{'}} \right\} = W({x_{p2}})
\end{array}.
\end{equation}

Next we embed the high-frequency information of the trigger image, i.e., $HP_{1,j}$ and $HP_{1,j}^{'}$ ($j=$ 1,2,3), into different high-frequency parts of the clean image $x_c$. Formally, we have: 
\begin{equation}
	\left\{ \begin{array}{l}
		H_{3,j}^{'} = {H_{3,j}} \times \alpha  + HP_{1,j}^{'} \times \left( {1 - \alpha } \right)\\
		H_{2,j}^{'} = {H_{2,j}} \times \beta  + HP_{1,j} \times \left( {1 - \beta } \right)
	\end{array}, \right.j=1,2,3,
\end{equation}
where $\alpha$ and $\beta$ indicate the embedding intensity. With $H_{3,j}^{'}$ and $H_{2,j}^{'}$, initial poisoned image $P_i$ can be derived using the IDWT as:
\begin{equation}
L_i^{'} = {W^ - }\left( {L_{i + 1}^{'},H_{i + 1,1}^{'},H_{i + 1,2}^{'},H_{i + 1,3}^{'}} \right), i=0,1,2,
\end{equation}
where $L_{3}^{'}$ equals $L_3$, $H_{1,j}^{'}$ equals $H_{1,j}$ and the final low-frequency component $L_0^{'}$ is exactly the generated poisoned image in Step 1, which is denoted as $P_i$. Considering that the high-frequency information is stealthier than the low-frequency information, the above method can generate an almost invisible backdoor image in the spatial domain (actually it depends on the intensity of $\alpha$ and $\beta$, we will discuss the effect of $\alpha$ and $\beta$ on the visual effect in the ablation experiments). In what follows, we aim to answer the question on how to craft a poisoned image that is also invisible in the frequency domain. 

\noindent\textbf{Step 2: Frequency Domain Smoothing.} \quad Previous studies~\cite{moreland2003fft,govindaraju2008high} have shown that the phase spectrum of an image after FFT retains information about the edges and the overall structure of the image, which captures high-level semantic information. Meanwhile, the amplitude spectrum obtains the underlying semantic information, preserving the frequency information~\cite{feng2022fiba}. Besides, it has been observed in ~\cite{yang2020fda} that changes in the amplitude spectrum do not significantly affect the perception of high-level semantics. Moreover, since the result of FFT is in the complex domain, while the image we usually observe in the frequency domain is actually the amplitude spectrum, we choose a straightforward yet effective approach. Specifically, to ensure both the backdoor attack performance and the amplitude spectrum's invisibility, we directly swap the amplitude spectrums of the $P_i$ and $x_c$ after FFT. Formally, as shown in Step 2 of Figure \ref{schematic}, let ${F^S}\left(  \cdot  \right)$ and ${F^P}\left(  \cdot  \right)$ be the amplitude and phase components of FFT results, the amplitude and phase spectra of $P_i$ and ${x_c}$ after FFT are obtained as:
\begin{equation}
	\left\{ \begin{array}{l}
		{S^c} = {F^S}\left( {{x_c}} \right),{P^c} = {F^P}\left( {{x_c}} \right)\\
		{S^p} = {F^S}\left( {P_i} \right),{P^p} = {F^P}\left( {P_i} \right)
	\end{array} \right..
\end{equation}

Then the smoothing poisoned image $P_{mi}$ is calculated by the amplitude spectrum of $x_c$ and the phase spectrum of $P_i$. Formally, 
\begin{equation}
	P_{mi} = {F^ - }\left( {{S^c},{P^p}} \right),
\end{equation}
where $F^-(\cdot)$ represents the inverse FFT. We would like to note that the FFT-based smoothing is conducted in the complex domain. Thus, despite the spectrogram of the image after FFT-based smoothing being theoretically hidden, the image is still perceivable in the real domain (see the ablation experiment for a visual demonstration). Moreover, the poisoned image may be detected by the DCT-based defense, which is unacceptable even though the detection probability is low.

To address this issue, next we incorporate the DCT, which is a special case of FFT in real domain, to fuse $P_{mi}$ and $x_c$. Due to the linear property of DCT, fusing the two images after only one DCT and then inversing the fused result is equivalent to directly fusing the two images, which can not achieve the purpose of deep smoothing. Therefore, we apply two DCTs on $P_{mi}$ and $x_c$ to achieve deeper information fusion. Let $D$ be the DCT while ${D^ - }$ be the inverse DCT (IDCT), we obtain the deep information in DCT domain as follows:
\begin{equation}
	D_c^{k} = D\left( {{D_c^{k-1}}} \right) , D_p^{k} = D\left( {{	D_p^{k-1}}} \right),\quad k = 1,2,
\end{equation}
where $k$ stands for the $k$-th DCT, $D_c^0$ equals $x_c$, and $D_p^0$ equals $P_{mi}$. As shown in Figure \ref{schematic}, the DCT smoothing is then implemented according to:
\begin{equation}
	D_p^{k - 1} = {D^ - }\left[ {D_p^k \times \lambda + D_c^k \times (1 - \lambda)} \right],\quad k=1,2,
\end{equation}
where $\lambda$ indicates the fusing intensity. After two steps of IDCT, we obtain the new $D_p^{0}$, which is exactly the intermediate poisoned image (denoted as $P_m$) generated in Step 2.

\noindent\textbf{Step 3: Random Trigger Masking.} \quad To ensure that the victim model learns the backdoor scattered over the entire image while well preserving the desirable attack stealthiness in both the spatial and frequency domains, we propose to randomly mask the trigger image. As shown in Step 3 of Figure \ref{schematic}, we first obtain the trigger embedded in $P_m$ by subtracting $x_c$ from $P_m$. Then, randomly mask it. Finally, we again embed the trigger pattern after masking into the clean image, yielding the final poisoned image $P_f$. 

\noindent\textbf{Attack Strategy Design.} \quad We further devise an attack strategy for DUBA to enhance both attack performance and stealthiness. Specifically, in the training phase, we adopt a weak trigger pattern via two operations, i.e., shrinking the values of $\alpha$ and $\beta$ as much as possible and masking more pixel points of the trigger image in Step 3. We employ a strong trigger pattern in the inference phase, which means that we make $\alpha$ and $\beta$ as large as possible while ensuring the triggers' invisibilities and masking fewer pixel points in Step 3. 

Moreover, considering that the triggers can be visible when the pixel values of points in the clean image are close to 0 or 255, we further mask the corresponding regions in the trigger image. In particular, in the training phase, we appropriately expand such regions for better attack stealthiness. 

\section{EXPERIMENTS}

\subsection{Experimental Settings}

\noindent\textbf{Datasets.}  \quad To evaluate the performance of DUBA on different tasks, we conduct experiments on four different datasets: 1) Cifar10~\cite{krizhevsky2009learning}, 2) Gtsrb~\cite{stallkamp2012man}, 3) ImageNet~\cite{deng2009imagenet}, and 4) Fer2013~\cite{goodfellow2013challenges}. Cifar10 and ImageNet are object classification datasets that include horses, aircraft, and other objects. Fer2013 is the face expression recognition dataset, while Gtsrb is the traffic signal recognition dataset. We present the details of these datasets in Table \ref{dataformation}. Note that the ImageNet is too large and we only use a subset of it. 
\begin{table}[]
    \small
    \caption{Dataset information.}
	\begin{tabular}{cccc}
		\toprule
 \textbf{Datasets}  & \textbf{Training/Testing Size}  & \textbf{Lables Size} & \textbf{Image Size}      \\ \midrule
		 Cifar10  & 50000/10000 & 10     & 32×32×3   \\
		 Gtsrb    & 39209/12603 & 43     & 64×64×3   \\
		 ImageNet & 48000/12000 & 100    & 224×224×3 \\
		 Fer2013 & 28708/3589  & 7      & 64×64×3   \\ \bottomrule
	\end{tabular}
	\centering
 \label{dataformation}
\end{table}


\begin{table*}[]\centering
	\caption{Attack effectiveness in terms of BA (\%) and ASR (\%). }
	\begin{tabular}{cc|cccccccccc}
		\toprule
		\multicolumn{1}{c}{\multirow{2}{*}{Models}}     & DataSet{$\rightarrow$} & \multicolumn{2}{c}{Cifar10}                      & \multicolumn{2}{c}{Gtsrb}                        & \multicolumn{2}{c}{ImageNet}                     & \multicolumn{2}{c}{Fer2013}                      \\
		\multicolumn{1}{c}{}                            & Methods{$\downarrow$}  & \multicolumn{1}{c}{BA (\%)} & \multicolumn{1}{c}{ASR (\%)} & \multicolumn{1}{c}{BA (\%)} & \multicolumn{1}{c}{ASR (\%)} & \multicolumn{1}{c}{BA (\%)} & \multicolumn{1}{c}{ASR (\%)} & \multicolumn{1}{c}{BA (\%)} & \multicolumn{1}{c}{ASR (\%)} \\ 		\midrule
		\multicolumn{1}{c|}{\multirow{6}{*}{ResNet18}}  & Clean   & 91.91                &                         & 99.32                &                         & 92.12                &                         & 63.22                &                         \\
		\multicolumn{1}{c|}{}                           & BadNets & 91.22                & 99.76                 & 99.14                & 99.62                 & 91.56                & \textbf{99.90}        & 62.72                & \textbf{99.99}        \\
		\multicolumn{1}{c|}{}                           & Blend   & 91.35                & 99.97                 & 99.06              & 99.72                 & 91.28                & 98.11                 & 63.09                & 99.72                 \\
		\multicolumn{1}{c|}{}                           & WaNet   & 91.25                & 99.78                 & 99.07                & 99.81                  &91.62                         &99.13                        & 62.41                & 99.01                 \\
		\multicolumn{1}{c|}{}                           & FIBA    & 91.08                & 99.26                 & 99.22                & 98.91                 & 92.02                & 98.96                 & 63.08                & 99.82                 \\

        \multicolumn{1}{c|}{}                       & \cellcolor{gray!18}DUBA    & \cellcolor{gray!18}91.55                & \cellcolor{gray!18}\textbf{99.98}        & \cellcolor{gray!18}99.21                & \cellcolor{gray!18}\textbf{99.92}        & \cellcolor{gray!18}91.55                & \cellcolor{gray!18}99.24                 & \cellcolor{gray!18}62.65                & \cellcolor{gray!18}99.89                 \\ \hline
		\multicolumn{1}{l|}{\multirow{6}{*}{RepVGG}}    & Clean   & 91.23                &                         & 99.46                &                         & 93.25                &                         & 61.59                &                         \\
		\multicolumn{1}{l|}{}                           & BadNets & 91.08                & 99.94                 & 99.42                & 99.87                 & 92.88                & \textbf{99.89}        & 61.52                & \textbf{99.92}        \\
		\multicolumn{1}{l|}{}                           & Blend   & 91.11                & 95.74                 & 99.13                & 99.56                 & 92.76                & 97.02                 & 60.55                & 99.90                 \\
		\multicolumn{1}{l|}{}                           & WaNet   & 91.11                & 99.51                 &99.17                        &99.28                     &92.36                      &99.17                        &61.28                      &99.72                        \\
		\multicolumn{1}{l|}{}                           & FIBA    & 90.08                & 99.11                 & 99.29                & 99.16                 & 92.72                & 98.94                 & 61.50                & 99.86                 \\

	\multicolumn{1}{l|}{}                           & \cellcolor{gray!18}DUBA   & \cellcolor{gray!18}91.18                & \cellcolor{gray!18}\textbf{99.98}        & \cellcolor{gray!18}99.41                & \cellcolor{gray!18}\textbf{99.89}        & \cellcolor{gray!18}92.52                & \cellcolor{gray!18}99.02                 & \cellcolor{gray!18}60.99                & \cellcolor{gray!18}99.91                 \\ \hline
		\multicolumn{1}{l|}{\multirow{6}{*}{Conformer}} & Clean   & 92.92                &                         & 99.50                &                         & 93.28                &                         & 63.88                &                         \\
		\multicolumn{1}{l|}{}                           & BadNets & 92.75                & \textbf{99.54}        & 99.39                & \textbf{99.79}        & 92.82                & \textbf{99.77}        & 63.70                & 99.89                 \\
		\multicolumn{1}{l|}{}                           & Blend   & 92.51                & 98.92                 & 99.45                & 99.06                 & 93.18                & 98.27                 & 63.72                & 99.84                 \\
		\multicolumn{1}{l|}{}                           & WaNet   & 92.36                   & 99.03                    &99.26                        &98.56                         &93.06                         & 99.01                         & 63.62                        & 99.75                         \\
		\multicolumn{1}{l|}{}                           & FIBA    & 92.08                & 98.93                 & 99.27                & 98.69                 &93.18                         &98.81                         & 62.51                & 99.82                 \\

		\multicolumn{1}{l|}{}                           & \cellcolor{gray!18}DUBA    & \cellcolor{gray!18}92.44                & \cellcolor{gray!18}99.12               & \cellcolor{gray!18}99.39                & \cellcolor{gray!18}99.57                 & \cellcolor{gray!18}93.11                & \cellcolor{gray!18}98.96                 & \cellcolor{gray!18}63.21                & \cellcolor{gray!18}\textbf{99.92}        \\ 		\bottomrule
	\end{tabular}

	\label{attack effectiness}
	\centering
\end{table*}
\noindent\textbf{Models.} \quad We conduct experiments on three models: ResNet18~\cite{he2016deep}, RepVGG~\cite{ding2021repvgg}, and Conformer~\cite{peng2021conformer}. ResNet18 is a classic classification model. RepVGG is the latest VGG model. Conformer is the latest transformer model for image classification.

\noindent\textbf{Baseline Backdoor Attacks.} \quad We compare DUBA with BadNets~\cite{gu2017badnets}, Blend~\cite{chen2017targeted}, WaNet~\cite{nguyen2021wanet}, and FIBA~\cite{feng2022fiba}. BadNets and Blend are representational visible backdoor attacks. WaNet is the latest invisible backdoor attack in the spatial domain while FIBA is the latest backdoor attack proposed from the frequency perspective.

\noindent\textbf{Evaluation Metrics.} \quad We evaluate DUBA and compare with baselines from two perspectives, i.e., attack performance and attack stealthiness. For attack performance evaluation, we employ the attack success rate (ASR), which is defined as the proportion of poisoned examples that are misclassified as the target label among all poisoned examples used for testing. Additionally, we utilize the benign accuracy (BA) to characterize the model's performance on clean testing data. For attack stealthiness evaluation, we use the following similarity metrics: peak signal-to-noise ratio (PSNR)~\cite{tanchenko2014visual}, structural similarity (SSIM)~\cite{hore2010image} and learned perceptual image patch similarity (LPIPS)~\cite{zhang2018unreasonable}. There are some correlations among these three metrics. Generally speaking, increased PSNR and SSIM indicate improved image steganography but decreased LPIPS.

\noindent\textbf{Implementation Details.} \quad  In our experiments, we randomly select an image with a dog's ear as the initial trigger image. In the training phase, we set both $\alpha$ and $\beta$ to 0.4 and mask the regions in the trigger image where the corresponding pixel points of the clean image are lower than 30 and more than 220. In the attack phase, we set $\alpha$ and $\beta$ both to 0.6, $\lambda$ to 0.7 and mask the regions in the trigger image which corresponds to pixel points lower than 5 or more than 245 in the clean image. We employ the SGD optimizer to train the victim model for 200 periods. The learning rate is set to 0.01 with a decay factor of 0.1 and decay periods of 50, 100, 150. The batch size is configured as 64. Following other studies~\cite{li2021invisible,feng2022fiba}, all attack settings are set to All-to-One attacks, which is sufficient for evaluating attack effectiveness. The defense experiments are all conducted on RepVgg model.

\subsection{Attack Performance Evaluation}
\noindent\textbf{Attack Effectiveness.} \quad We evaluate the effectiveness of different backdoor attacks with ASR and BA. The relevant results are summarized in Table \ref{attack effectiness}, which shows that our proposed DUBA achieves higher or comparable ASRs under most datasets and models. But in some cases such as experiments under ImageNet, the ASRs of DUBA are slightly lower than BadNets. Considering that the crafted trigger by DUBA is invisible in both spatial and frequency domains (which will be validated next), such a result is acceptable. Besides, DUBA only incurs negligible loss (lower than 1\%) of BA compared with the clean benchmark. The above results show that DUBA achieves desirable attack effectiveness. 

\begin{figure}[]
	\centering
	\includegraphics[height=2.75cm,width=8cm]{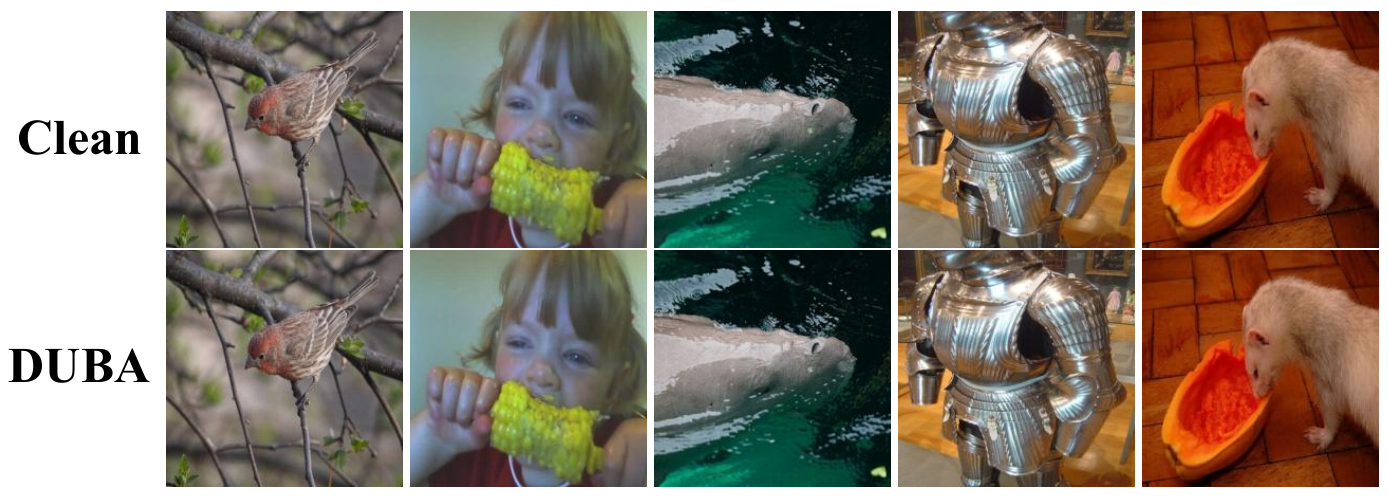}
	\caption{Visual comparison of clean images and images poisoned by DUBA.}
\label{com}
\end{figure}

\noindent\textbf{Attack Stealthiness.} \quad Now we examine the stealthiness of different backdoor attacks. Figure \ref{visuall} shows the poisoned images of different methods and Figure \ref{com} provides more visual comparison between clean images and images poisoned by DUBA. Compared with other methods, DUBA achieves the best invisibility in both the spatial and frequency domains. The backdoor generated by DUBA is visually invisible in the spatial domain and the residual image in the frequency domain is also close to pure black image, indicating that the poisoned image in the frequency domain is very similar to the clean image. In Table \ref{attack stealthiness}, the visual outcomes of various methods are quantified. The PSNR and LPIPS of DUBA is the best in most cases. Although the SSIM of DUBA is slightly lower than BadNets, it is also close to 1 and higher than most methods. It can be validated from Figure \ref{visuall} that BadNets has the worst stealthiness due to its obvious square trigger in the corner. In summary, DUBA achieved the best stealthy results with comprehensive visual perception and different metrics.
\begin{table*}[]
\centering
\small
\caption{Stealthiness of different attacks in the spatial domain. }
\begin{tabular}{c|p{6mm}p{6mm}p{6.6mm}|p{6mm}p{6mm}p{6.6mm}|p{6mm}p{6mm}p{6.6mm}|p{6mm}p{6mm}p{6.6mm}|p{6mm}p{6mm}p{6.5mm}}
\toprule
\textbf{Methods}  & \multicolumn{3}{c|}{\textbf{BadNets}} & \multicolumn{3}{c|}{\textbf{Blend}} & \multicolumn{3}{c|}{\textbf{WaNet}} & \multicolumn{3}{c|}{\textbf{FIBA}} & \multicolumn{3}{c}{\textbf{DUBA}}        \\
\textbf{METRIC}   & PSNR{$\uparrow$}     & SSIM{$\uparrow$}              & LPIPS{$\downarrow$}     & PSNR{$\uparrow$}        & SSIM{$\uparrow$}       & LPIPS{$\downarrow$}      & PSNR{$\uparrow$}        & SSIM{$\uparrow$}       & LPIPS{$\downarrow$}      & PSNR{$\uparrow$}       & SSIM{$\uparrow$}       & LPIPS{$\downarrow$}      & PSNR{$\uparrow$}            & SSIM{$\uparrow$}   & LPIPS{$\downarrow$}           \\ \midrule
\textbf{Cifar10}  & 30.19   & \textbf{0.989}   & 0.0096   & 18.96      & 0.862     & 0.1219     & 33.98      & 0.966     & 0.0090     & 29.46     & 0.962     & 0.0115     & \textbf{37.98} & 0.977 & \textbf{0.0081} \\
\textbf{Gtsrb}    & 34.51   & \textbf{0.983}   & 0.0351   & 19.81      & 0.887     & 0.1661     & 30.51      & 0.960     & 0.0588     & 30.17     & 0.975     & 0.0622     & \textbf{36.01} & 0.972 & \textbf{0.0201} \\
\textbf{ImageNet} & 25.14   & \textbf{0.991}   & 0.0079   & 20.56      & 0.896     & 0.1831     & 29.65      & 0.951     & 0.0932     & 30.56     & 0.969     & 0.0764     & \textbf{36.22} & 0.976 & \textbf{0.0072} \\
\textbf{Fer2013}  & 30.28   & \textbf{0.994}   & 0.0087   & 16.56      & 0.824     & 0.1957     & 31.26      & 0.962     & 0.0081     & 33.46     & 0.994     & \textbf{0.0079}     & \textbf{36.14} & 0.962 & 0.0083 \\ \bottomrule

\end{tabular}

\centering
\label{attack stealthiness}
\end{table*}
\subsection{Robustness to Defenses}
In this subsection, we test DUBA against five state-of-the-art defenses, including GradCam~\cite{selvaraju2017grad}, Neural Cleanse~\cite{wang2019neural},  STRIP~\cite{gao2019strip}, Fine-Prunning~\cite{liu2018fine}, and FTD~\cite{zeng2021rethinking}.

\begin{figure}[]
	
	\centering
	\includegraphics[height=4.7cm,width=8cm]{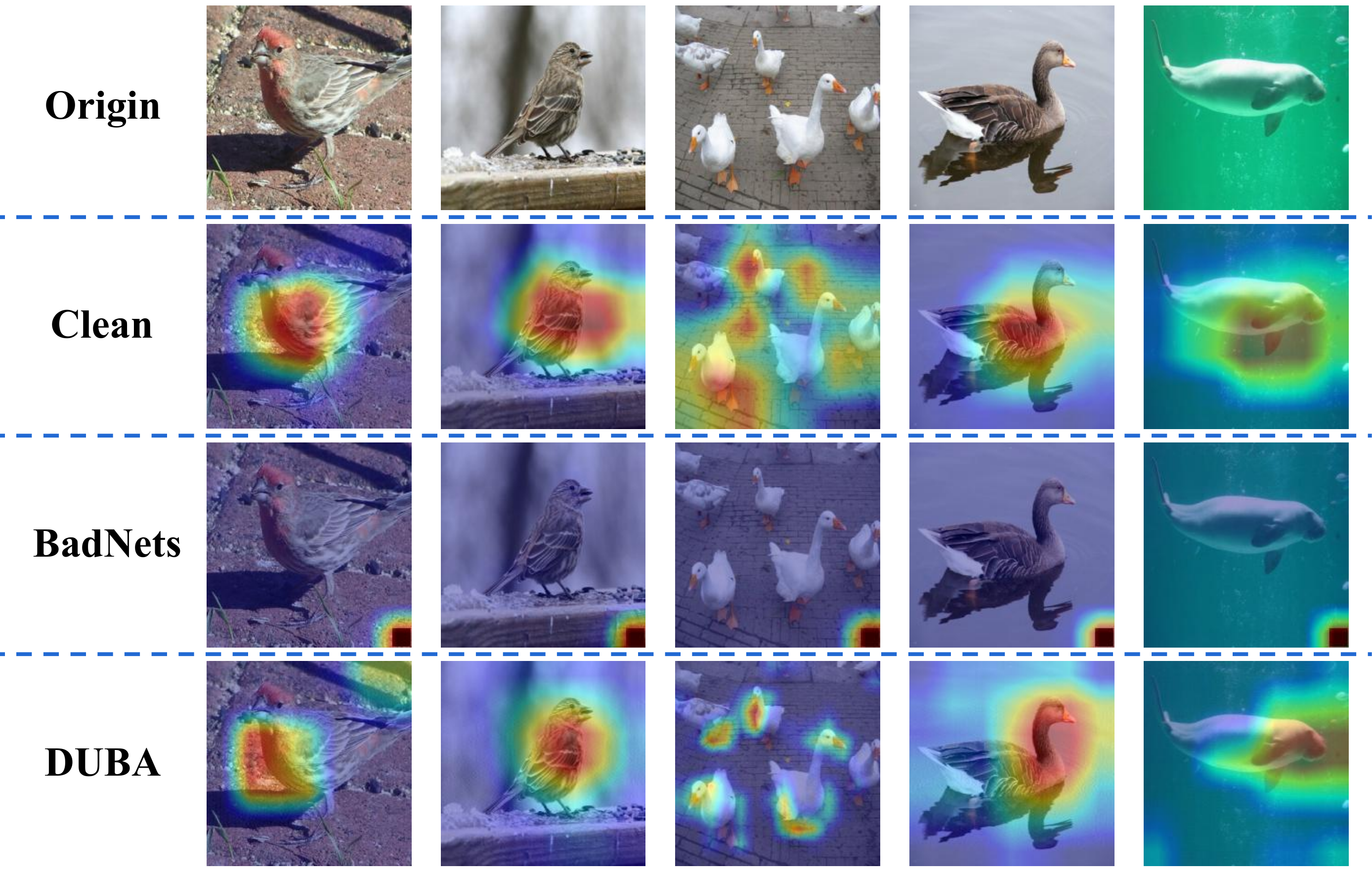}
	\caption{The GradCam of different samples. If GradCam successfully defends against a backdoor attack like badnets, it will lock onto the trigger. GradCam for DUBA is similar to clean images, and even locks onto the object more than clean images. }
	\label{gradcam}
\end{figure}

\noindent\textbf{Robustness to GradCam.}  \quad The GradCam-based defense method uses the saliency map to analyze the model's decision process. Specifically, given an input sample to the model, GradCam yields the model's heat value. For a clean image, GradCam will focus on the object. As shown in Figure \ref{gradcam}, for small triggers such as BadNets, the heat map locks the highest heat value on the trigger, resulting in an abnormal heat map. The results show that GradCam for DUBA is similar to clean images, and even locks onto the object more than clean images. This indicates that GradCam fails to detect DUBA. 

\noindent\textbf{Robustness to Neural Cleanse.} \quad Neural Cleanse reconstructs the trigger for each class label, and then checks whether there exists a class with a significantly smaller reverse-engineered trigger, which will be treated as a poisoned sample. Specifically, this method quantifies the deviations of reverse-engineered triggers based on their sizes using the anomaly index and considers models with an anomaly index greater than 2 as poisoned models. Table \ref{anomaly index} shows that the anomaly index of DUBA is only 1.22, which is smaller than those of the baseline methods. This validates that our proposed DUBA can effectively circumvent Neural Cleanse. 
\begin{table}[]
	\caption{Anomaly index values for different attacks.}
	\begin{tabular}{c|p{5mm}p{8mm}p{6mm}p{6mm}p{4mm}p{6.5mm}}
		\toprule
		\textbf{Methods}      & Clean & BadNets & Blend & WaNet & FIBA & DUBA            \\ \midrule
		\textbf{Anomaly Index{$\downarrow$}} & 0.76           & 4.28             & 3.26           &       2.32         & 1.92          & \multicolumn{1}{c}{\textbf{1.22}} \\ 		\bottomrule
	\end{tabular}
	\label{anomaly index}
	\centering
\end{table}

\begin{figure}[b]
	
	\centering
	\includegraphics[height=3.9cm]{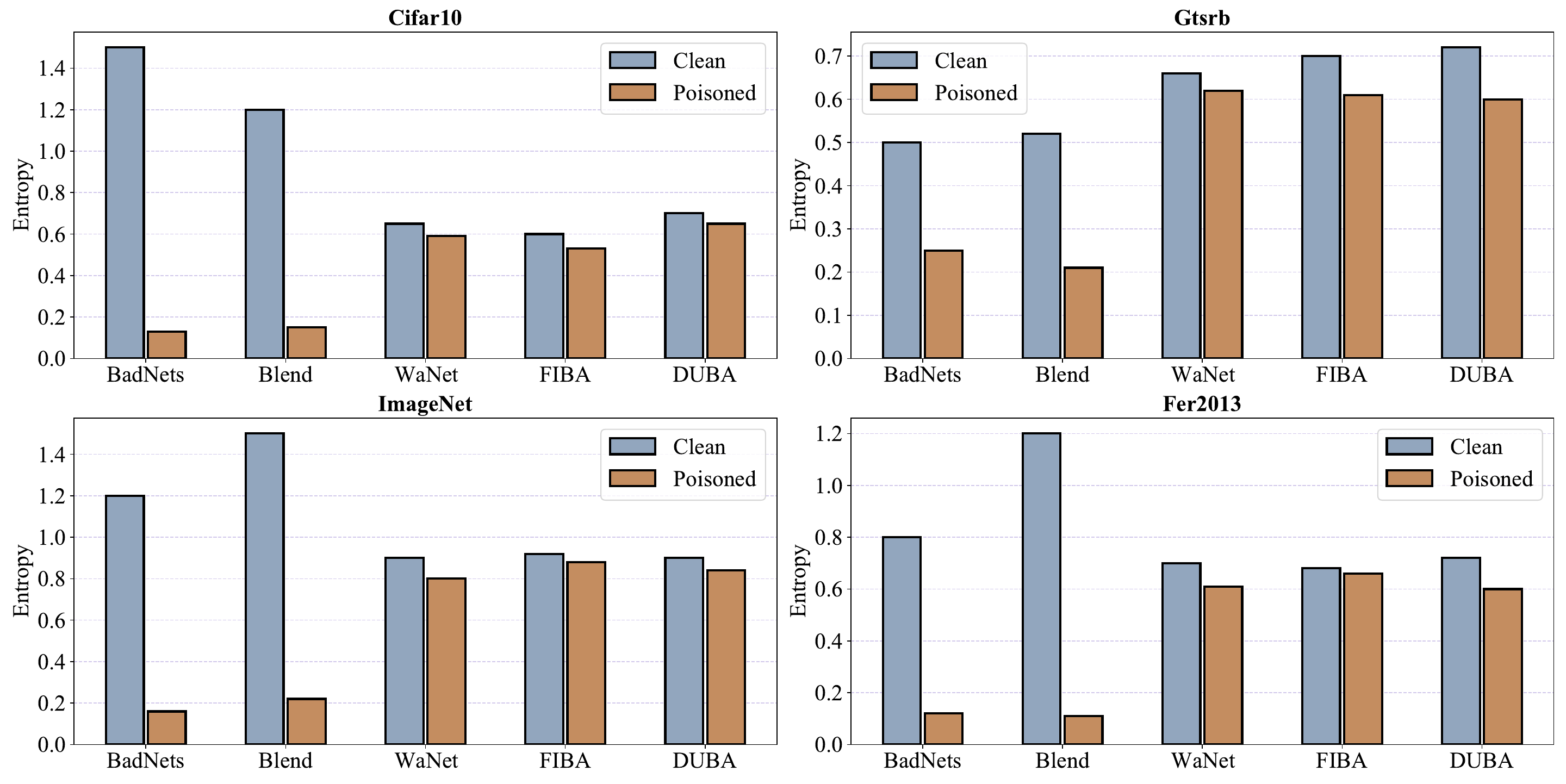}
	\caption{Entropy values of STRIP on different datasets. The entropy values of poisoned images on DUBA are very close to those of clean images.}
 \label{entropy}
\end{figure}

\noindent\textbf{Robustness to STRIP.}  \quad STRIP determines whether a model is poisoned or not by superimposing input images to observe the consistency of predicted classes. Specifically, the entropy values is used to quantify the level of consistency. Models with an average entropy value lower than 0.2 and the clean results are classified as poisoned models. Figure \ref{entropy} shows the entropy values of poisoned images by different methods on different datasets and the corresponding clean results. All the entropy values of DUBA are larger than 0.2 and very close to
that of clean images, which are significantly better than BadNets and Blend. DUBA also achieves comparable entropy values to WaNet and FIBA. Thus, DUBA can effectively bypass STRIP.

\begin{figure}[b]
	\centering	\includegraphics[height=4.3cm,width=8.45cm]{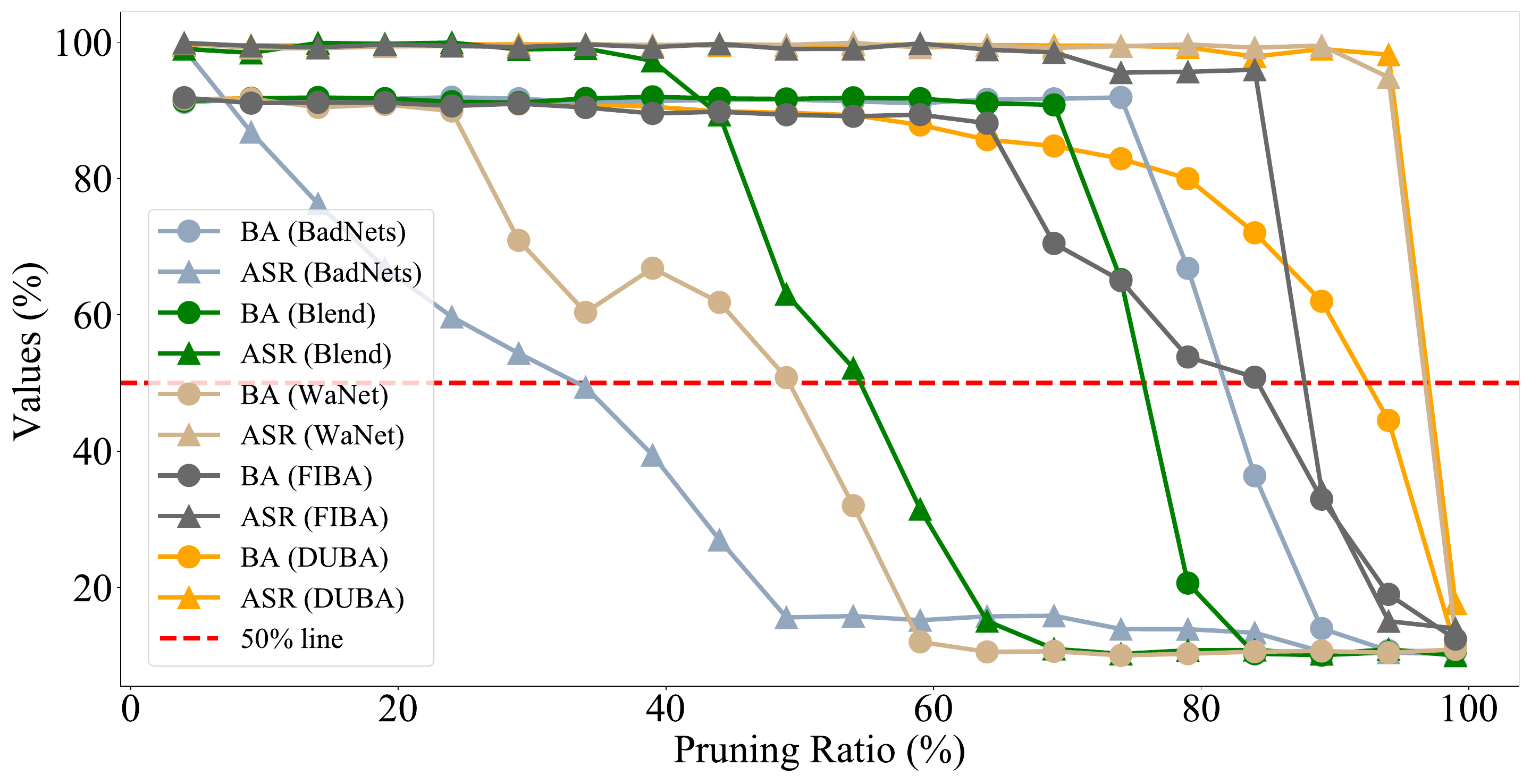}
	\caption{The results of Fine-Pruning on different methods. The ASR of DUBA is the latest to start declining.}
 \label{fine}
\end{figure}

\begin{figure}[b]
	\centering	\includegraphics[height=4.3cm,width=8.45cm]{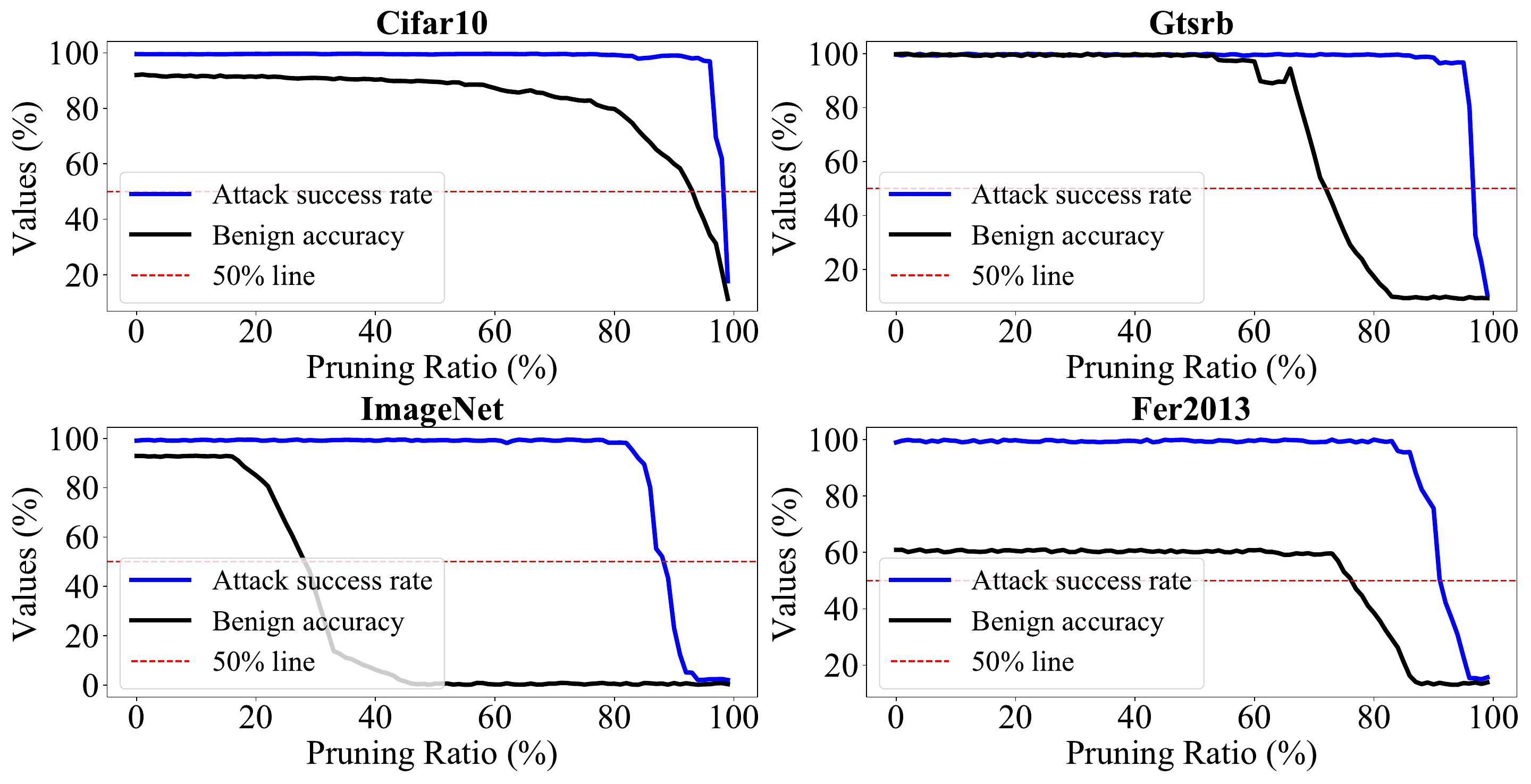}
	\caption{The results of Fine-Pruning on different datasets. In four cases, the BA falls below 50\% before the ASR.}
 \label{fine2}
\end{figure}

\noindent\textbf{Robustness to Fine-Pruning.} \quad Fine-pruning assumes that the backdoor behavior of the model is related to the dormant neurons in the model. By simply clipping these neurons, a
clean model will be obtained. Specifically, it records the activation values of clean samples passing through each neuron and considers the neuron with the smallest activation value as the most dormant neuron. The neurons are gradually pruned from small to large according to neuronal activation values. Usually, the attack is considered successful if the BA of clean images drops below 50\% before (in terms of pruning ratio) the ASR of poisoned images. Figure \ref{fine} shows the results of different methods on Cifar10. Among all the methods, the ASR of DUBA is the last to decline. In particular, when the ASR of DUBA starts to decrease, the pruning ratio almost reaches 96\%. Figure \ref{fine2} provides more detailed results on DUBA regarding the four datasets, which shows that all the BAs decrease to 50\% before ASRs. Thus, we can conclude that DUBA remains effective against network pruning-based defenses. 

\begin{table}[b]
\caption{Detection rates of FTD for different attack methods. }	
	\begin{tabular}{c|p{9mm}p{6.5mm}p{6mm}p{5mm}p{5.5mm}}
		\toprule
		
		\textbf{Attack Methods}       & BadNets & Blend & WaNet & FIBA & DUBA             \\ \midrule
		\textbf{Detection Rate (\%)} & 91.72       & 96.41       & 68.95        & 58.2       & \multicolumn{1}{p{8mm}}{\textbf{49.96}} \\ 		\bottomrule
	\end{tabular}

	\centering
        \label{FTD}
\end{table}

\noindent\textbf{Robustness to FTD.}  \quad FTD detects whether an image has high-frequency artifacts, which is regarded as poisoned image. It trains a DNN model that can classify poisoned images and clean images after DCT. Table \ref{FTD} presents that the detection rate of FTD with respect to DUBA is below 50\%, implying that DUBA can effectively bypass FTD. This is because the trigger image is smoothed twice in the frequency domain. Note that FIBA also presents a low detection ratio (higher than DUBA) as it uses a low-frequency trigger. 

\subsection{Ablation Studies}
In this section, we conduct ablation experiments to study the impact of some important parts on DUBA. Experiments are conducted on Cifar10 for training RepVgg.

\noindent\textbf{High-Frequency Embedding Rate.} \quad We first examine the effect of $\alpha$ and $\beta$ (the embedding ratio of the trigger image) on the ASR. Table \ref{abab} shows that DUBA yields a lower ASR when the embedding ratio during training is small. This can be attributed to the inability of the model to learn the complete backdoor or the large differences in the embedding amount between the training and inference phases. When the embedding ratio rises in both the training and attacking phases, DUBA achieves higher ASRs. 
\begin{table}[]
	\centering
 	\caption{Impact of high-frequency embedding ratio on ASR. }
	\begin{tabular}{clll|l}
		
		\toprule
		\multicolumn{2}{c}{\textbf{Train}}                       & 
		\multicolumn{2}{c|}{\textbf{Attack}}        & \multirow{2}{*}{\textbf{ASR (\%)}} \\ 
		\textbf{$\alpha$}              & \multicolumn{1}{c}{\textbf{$\beta$}} & \multicolumn{1}{c}{\textbf{$\alpha$}} & \textbf{$\beta$} &                               \\ \midrule
		0.2                     & 0.2                            & 0.2                            & 0.2        & 85.62                       \\
		0.2                     & 0.2                            & 0.6                            & 0.6        & 87.92                       \\
		0.4                     & 0.4                            & 0.6                            & 0.6        & 99.98                       \\
		0.4                     & 0.4                            & 0.8                            & 0.8        & 99.99                       \\
		\multicolumn{1}{l}{0.6} & 0.6                            & 0.6                            & 0.6        & 99.22\\
		
		\bottomrule
		
	\end{tabular}
        \label{abab}
	\centering
\end{table}

\noindent\textbf{DCT Smoothing Parameters.} \quad We also investigate the effect of the DCT smoothing parameter on the ASR. Intuitively, when $\lambda$ decreases, the poisoned images will be closer to clean ones. Thus, the ASR is compromised while the attack stealthiness is enhanced. Table \ref{dctdct} shows that the ASR increases with $\lambda$, which is consistent with the intuition. 
\begin{table}[]
	\caption{Impact of DCT smoothing parameter $\lambda$ on ASR.}
	\begin{tabular}{c|cccccc}
		\toprule
		\textbf{$\lambda$} & 0.1& \multicolumn{1}{c}{0.2}  & \multicolumn{1}{c}{0.3} & \multicolumn{1}{c}{0.5} & \multicolumn{1}{c}{0.7} & \multicolumn{1}{c}{0.8} \\ \midrule
		\textbf{ASR(\%)} & 20 & 52.36 & 70.11                 & 95.42                 & 99.98                 & 99.98                 \\ 		\bottomrule
	\end{tabular}

	\label{dctdct}
	\centering
\end{table}

\noindent\textbf{Initial Trigger Selection.} \quad We explore the effect of different initial trigger images on DUBA. In addition to the image of dog's ear used as the initial trigger image in the above experiments, three other images in Cifar10, Gtsrb, and ImageNet are also tested. Table \ref{trigger selection} shows that there is no substantial association between the initial trigger and ASR.
\begin{table}[H]
	\caption{Impact of different initial trigger images on ASR.}
	\begin{tabular}{c|p{13mm}p{7.9mm}p{5.9mm}p{10.4mm}}
		\toprule
		\textbf{Initial Trigger Image} & Dog's ear & Cifar10 & Gtsrb & ImageNet \\ \midrule
		\textbf{ASR on Cifar10 (\%)}    & 99.98     & 99.23   & 99.57 & 99.62    \\ \bottomrule
	\end{tabular}

	\label{trigger selection}
	\centering
\end{table}

\noindent\textbf{The Necessity of Three 
Frequency Domain Transforms.} \quad In the following three subsections, we conduct ablation experiments to show the necessity of three frequency domain transforms.

\begin{figure}[]
	\centering
	\includegraphics[height=5.1cm,width=8.1cm]{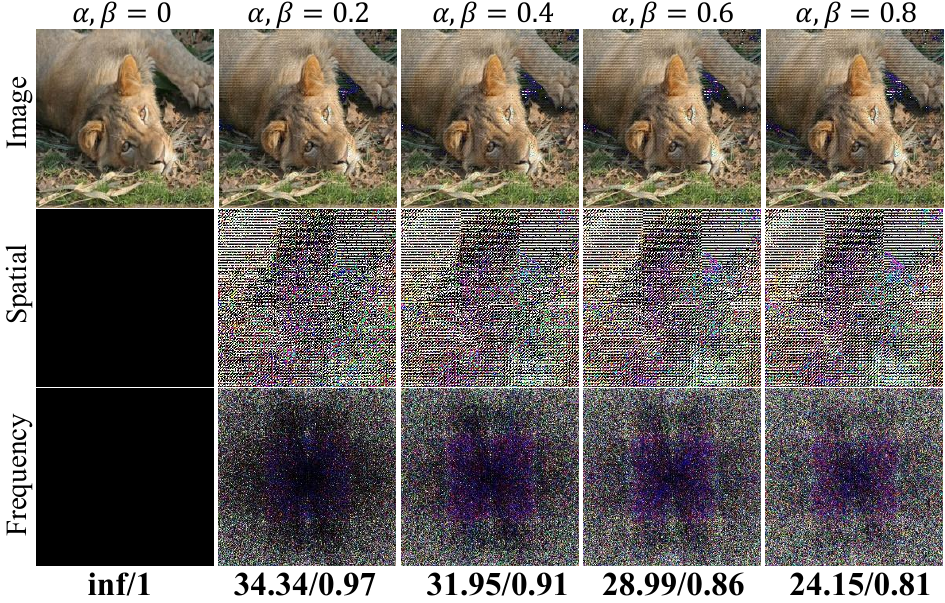}
	\caption{The visual results of using DWT only. From left to right, the embedding coefficient gradually rises from 0 to 0.8. The first row shows the poisoned images with different embedding coefficients (the first one is the clean picture). The second row shows the corresponding residual images in the spatial domain and the third row are the residual images in the frequency domain. The corresponding PSNR and SSIM values (PSNR/SSIM) are at the bottom of the figure.}
 \label{88}
\end{figure}

\noindent\textbf{Use DWT Only.} \quad We conduct experiments using only DWT and do not apply any subsequent smoothing steps to the output poisoned image, as shown in Step 1 of Figure \ref{schematic}. Figure \ref{88} visualizes the results. Although the PSNR and SSIM values between the poisoned and clean images are high enough when the embedding coefficients are small, the poisoned images inevitably have visible artifacts in the frequency domain, making it impossible to achieve dual stealth (similar to the previous single stealth backdoor study). This demonstrates the necessity of following smoothing operations.

\begin{figure}[]
	\centering
	\includegraphics[height=5cm,width=6.58cm]{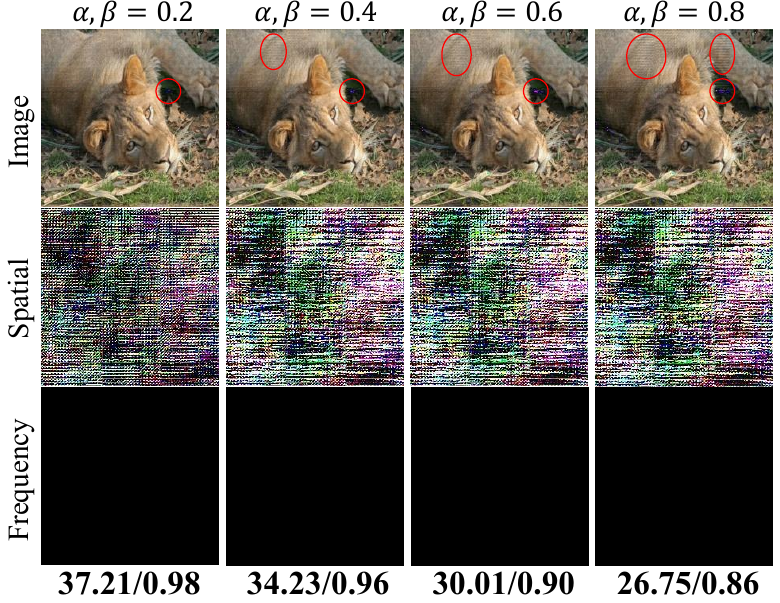}
	\caption{The visual results of using DWT and FFT. The red circles indicate the line-like artifacts.}
 \label{99}
\end{figure}

\noindent\textbf{Use DWT and FFT.} \quad We then conduct experiments where the poisoned image is only smoothed in the FFT domain. Figure \ref{99} shows that even when $\alpha$ and $\beta$ are large enough, the residual images in the frequency domain are pure black (i.e., the poisoned images are invisible in the frequency domain). Although the embedding ratio is small enough, the poisoned images (first row) still have some line-like artifacts in the spatial domain. Furthermore, when $\alpha$ and $\beta$ are set to 0.4, FTD can detect our attack with a probability of about 65\%, which is already lower than most attacks but is unacceptable. This demonstrates the necessity of the DCT-based smoothing operations.

\begin{figure}[]
	\centering
	\includegraphics[height=5cm,width=6.58cm]{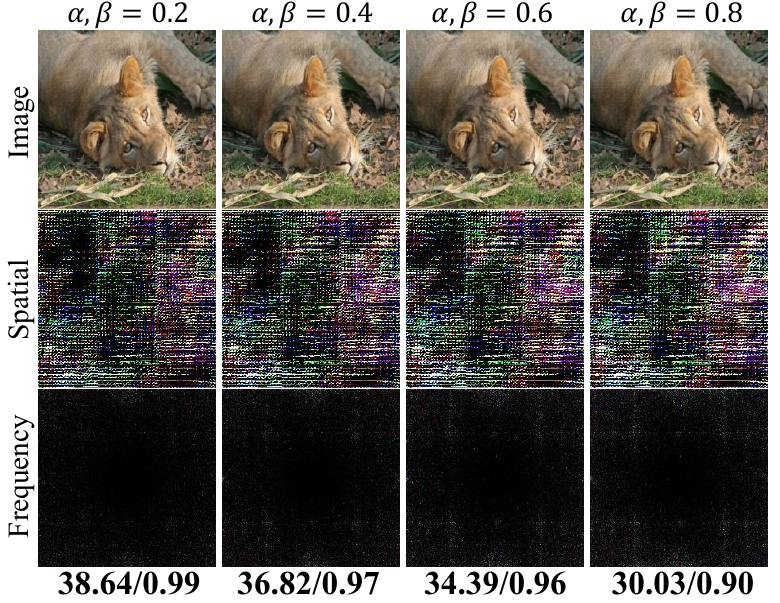}
	\caption{The visual results of using three transforms.}
 \label{1010}
\end{figure}

\noindent\textbf{Use DWT, FFT and DCT.} \quad According to Table \ref{dctdct}, we set $\lambda$ to 0.7. As shown in Figure \ref{1010}, after using the three transforms, the poisoned image is stealthy in both domains (both the PSNR and SSIM have been improved). Thus, the three frequency domain transforms are adopted in the proposed DUBA to achieve dual stealth.


\subsection{Summary of Experiments}
After the experimental comparsion, we can conclude that DUBA is substantially more stealthy in the spatial domain than other attacks, which is also the only attack that is simultaneously invisible in both the spatial and frequency domains. Furthermore, DUBA achieves remarkable ASRs that outperform other methods in most cases. It is validated that the five advanced defenses fail to detect DUBA. And also it shows that DUBA has stronger robustness than other methods. Although there are some specific cases, such as the robustness on STRIP of Fer2013, where DUBA performs slightly worse than WaNet or FIBA, in most
cases, especially in terms of the invisibility in frequency domain, DUBA significantly outperforms all the methods. Thus we can conclude that the proposed DUBA is effective, which outperforms the state-of-the-art backdoor attacks.

\section{Conclusion}\label{sec:conclusion}
In this paper, we showed that most backdoor attacks are visible in the frequency domain. In order to completely break the defense proposed from the frequency perspective while remaining stealthy in the spatial domain, we proposed a \textbf{DU}al stealthy \textbf{BA}ckdoor called DUBA that is invisible in both the spatial and frequency domains. To hide high-frequency backdoor information in both the spatial and frequency domains, we leveraged the benefits of different frequency domain transforms. A novel attack strategy was also devised in order to enhance the efficiency of DUBA. We conducted an extensive experimental evaluation of DUBA. The results corroborate its outstanding performance in terms of attack success rates and attack stealthiness.


\newpage
\bibliographystyle{ACM-Reference-Format}
\bibliography{sample-base}


\begin{thebibliography}{36}


\ifx \showCODEN    \undefined \def \showCODEN     #1{\unskip}     \fi
\ifx \showDOI      \undefined \def \showDOI       #1{#1}\fi
\ifx \showISBNx    \undefined \def \showISBNx     #1{\unskip}     \fi
\ifx \showISBNxiii \undefined \def \showISBNxiii  #1{\unskip}     \fi
\ifx \showISSN     \undefined \def \showISSN      #1{\unskip}     \fi
\ifx \showLCCN     \undefined \def \showLCCN      #1{\unskip}     \fi
\ifx \shownote     \undefined \def \shownote      #1{#1}          \fi
\ifx \showarticletitle \undefined \def \showarticletitle #1{#1}   \fi
\ifx \showURL      \undefined \def \showURL       {\relax}        \fi
\providecommand\bibfield[2]{#2}
\providecommand\bibinfo[2]{#2}
\providecommand\natexlab[1]{#1}
\providecommand\showeprint[2][]{arXiv:#2}

\bibitem[Chen et~al\mbox{.}(2017)]%
        {chen2017targeted}
\bibfield{author}{\bibinfo{person}{Xinyun Chen}, \bibinfo{person}{Chang Liu},
  \bibinfo{person}{Bo Li}, \bibinfo{person}{Kimberly Lu}, {and}
  \bibinfo{person}{Dawn Song}.} \bibinfo{year}{2017}\natexlab{}.
\newblock \showarticletitle{Targeted backdoor attacks on deep learning systems
  using data poisoning}.
\newblock \bibinfo{journal}{\emph{arXiv preprint arXiv:1712.05526}}
  (\bibinfo{year}{2017}).
\newblock


\bibitem[Chou et~al\mbox{.}(2020)]%
        {chou2020sentinet}
\bibfield{author}{\bibinfo{person}{Edward Chou}, \bibinfo{person}{Florian
  Tramer}, {and} \bibinfo{person}{Giancarlo Pellegrino}.}
  \bibinfo{year}{2020}\natexlab{}.
\newblock \showarticletitle{Sentinet: Detecting localized universal attacks
  against deep learning systems}. In \bibinfo{booktitle}{\emph{IEEE Security
  and Privacy Workshops}}. \bibinfo{pages}{48--54}.
\newblock


\bibitem[Cintra and Bayer(2011)]%
        {cintra2011dct}
\bibfield{author}{\bibinfo{person}{Renato~J Cintra} {and}
  \bibinfo{person}{F{\'a}bio~M Bayer}.} \bibinfo{year}{2011}\natexlab{}.
\newblock \showarticletitle{A DCT approximation for image compression}.
\newblock \bibinfo{journal}{\emph{IEEE Signal Processing Letters}}
  (\bibinfo{year}{2011}), \bibinfo{pages}{579--582}.
\newblock


\bibitem[Deng et~al\mbox{.}(2009)]%
        {deng2009imagenet}
\bibfield{author}{\bibinfo{person}{Jia Deng}, \bibinfo{person}{Wei Dong},
  \bibinfo{person}{Richard Socher}, \bibinfo{person}{Li-Jia Li},
  \bibinfo{person}{Kai Li}, {and} \bibinfo{person}{Li Fei-Fei}.}
  \bibinfo{year}{2009}\natexlab{}.
\newblock \showarticletitle{Imagenet: A large-scale hierarchical image
  database}. In \bibinfo{booktitle}{\emph{Proceedings of the IEEE/CVF
  Conference on Computer Vision and Pattern Recognition}}.
  \bibinfo{pages}{248--255}.
\newblock


\bibitem[Ding et~al\mbox{.}(2021)]%
        {ding2021repvgg}
\bibfield{author}{\bibinfo{person}{Xiaohan Ding}, \bibinfo{person}{Xiangyu
  Zhang}, \bibinfo{person}{Ningning Ma}, \bibinfo{person}{Jungong Han},
  \bibinfo{person}{Guiguang Ding}, {and} \bibinfo{person}{Jian Sun}.}
  \bibinfo{year}{2021}\natexlab{}.
\newblock \showarticletitle{Repvgg: Making vgg-style convnets great again}. In
  \bibinfo{booktitle}{\emph{Proceedings of the IEEE/CVF Conference on Computer
  Vision and Pattern Recognition}}. \bibinfo{pages}{13733--13742}.
\newblock


\bibitem[Doan et~al\mbox{.}(2021)]%
        {doan2021lira}
\bibfield{author}{\bibinfo{person}{Khoa Doan}, \bibinfo{person}{Yingjie Lao},
  \bibinfo{person}{Weijie Zhao}, {and} \bibinfo{person}{Ping Li}.}
  \bibinfo{year}{2021}\natexlab{}.
\newblock \showarticletitle{Lira: Learnable, imperceptible and robust backdoor
  attacks}. In \bibinfo{booktitle}{\emph{Proceedings of the IEEE/CVF
  International Conference on Computer Vision}}. \bibinfo{pages}{11966--11976}.
\newblock


\bibitem[Feng et~al\mbox{.}(2022)]%
        {feng2022fiba}
\bibfield{author}{\bibinfo{person}{Yu Feng}, \bibinfo{person}{Benteng Ma},
  \bibinfo{person}{Jing Zhang}, \bibinfo{person}{Shanshan Zhao},
  \bibinfo{person}{Yong Xia}, {and} \bibinfo{person}{Dacheng Tao}.}
  \bibinfo{year}{2022}\natexlab{}.
\newblock \showarticletitle{FIBA: Frequency-Injection based Backdoor Attack in
  Medical Image Analysis}. In \bibinfo{booktitle}{\emph{Proceedings of the
  IEEE/CVF Conference on Computer Vision and Pattern Recognition}}.
  \bibinfo{pages}{20876--20885}.
\newblock


\bibitem[Gao et~al\mbox{.}(2019)]%
        {gao2019strip}
\bibfield{author}{\bibinfo{person}{Yansong Gao}, \bibinfo{person}{Change Xu},
  \bibinfo{person}{Derui Wang}, \bibinfo{person}{Shiping Chen},
  \bibinfo{person}{Damith~C Ranasinghe}, {and} \bibinfo{person}{Surya Nepal}.}
  \bibinfo{year}{2019}\natexlab{}.
\newblock \showarticletitle{Strip: A defence against trojan attacks on deep
  neural networks}. In \bibinfo{booktitle}{\emph{Proceedings of the 35th Annual
  Computer Security Applications Conference}}. \bibinfo{pages}{113--125}.
\newblock


\bibitem[Goodfellow et~al\mbox{.}(2013)]%
        {goodfellow2013challenges}
\bibfield{author}{\bibinfo{person}{Ian~J Goodfellow}, \bibinfo{person}{Dumitru
  Erhan}, \bibinfo{person}{Pierre~Luc Carrier}, \bibinfo{person}{Aaron
  Courville}, \bibinfo{person}{Mehdi Mirza}, \bibinfo{person}{Ben Hamner},
  \bibinfo{person}{Will Cukierski}, \bibinfo{person}{Yichuan Tang},
  \bibinfo{person}{David Thaler}, \bibinfo{person}{Dong-Hyun Lee},
  {et~al\mbox{.}}} \bibinfo{year}{2013}\natexlab{}.
\newblock \showarticletitle{Challenges in representation learning: A report on
  three machine learning contests}. In \bibinfo{booktitle}{\emph{Proceedings of
  the International Conference on Neural Information Processing}}.
  \bibinfo{pages}{117--124}.
\newblock


\bibitem[Govindaraju et~al\mbox{.}(2008)]%
        {govindaraju2008high}
\bibfield{author}{\bibinfo{person}{Naga~K Govindaraju},
  \bibinfo{person}{Brandon Lloyd}, \bibinfo{person}{Yuri Dotsenko},
  \bibinfo{person}{Burton Smith}, {and} \bibinfo{person}{John Manferdelli}.}
  \bibinfo{year}{2008}\natexlab{}.
\newblock \showarticletitle{High performance discrete Fourier transforms on
  graphics processors}. In \bibinfo{booktitle}{\emph{Proceedings of the
  ACM/IEEE conference on Supercomputing}}. \bibinfo{pages}{1--12}.
\newblock


\bibitem[Gu et~al\mbox{.}(2017)]%
        {gu2017badnets}
\bibfield{author}{\bibinfo{person}{Tianyu Gu}, \bibinfo{person}{Brendan
  Dolan-Gavitt}, {and} \bibinfo{person}{Siddharth Garg}.}
  \bibinfo{year}{2017}\natexlab{}.
\newblock \showarticletitle{Badnets: Identifying vulnerabilities in the machine
  learning model supply chain}.
\newblock \bibinfo{journal}{\emph{arXiv preprint arXiv:1708.06733}}
  (\bibinfo{year}{2017}).
\newblock


\bibitem[He et~al\mbox{.}(2016)]%
        {he2016deep}
\bibfield{author}{\bibinfo{person}{Kaiming He}, \bibinfo{person}{Xiangyu
  Zhang}, \bibinfo{person}{Shaoqing Ren}, {and} \bibinfo{person}{Jian Sun}.}
  \bibinfo{year}{2016}\natexlab{}.
\newblock \showarticletitle{Deep residual learning for image recognition}. In
  \bibinfo{booktitle}{\emph{Proceedings of the IEEE/CVF Conference on Computer
  Vision and Pattern Recognition}}. \bibinfo{pages}{770--778}.
\newblock


\bibitem[Hore and Ziou(2010)]%
        {hore2010image}
\bibfield{author}{\bibinfo{person}{Alain Hore} {and} \bibinfo{person}{Djemel
  Ziou}.} \bibinfo{year}{2010}\natexlab{}.
\newblock \showarticletitle{Image quality metrics: PSNR vs. SSIM}. In
  \bibinfo{booktitle}{\emph{Proceedings of International Conference on Pattern
  Recognition}}. \bibinfo{pages}{2366--2369}.
\newblock


\bibitem[Huang et~al\mbox{.}(2020)]%
        {huang2020one}
\bibfield{author}{\bibinfo{person}{Shanjiaoyang Huang}, \bibinfo{person}{Weiqi
  Peng}, \bibinfo{person}{Zhiwei Jia}, {and} \bibinfo{person}{Zhuowen Tu}.}
  \bibinfo{year}{2020}\natexlab{}.
\newblock \showarticletitle{One-pixel signature: Characterizing cnn models for
  backdoor detection}. In \bibinfo{booktitle}{\emph{European Conference on
  Computer Vision}}. \bibinfo{pages}{326--341}.
\newblock


\bibitem[Jacob et~al\mbox{.}(2017)]%
        {fix}
\bibfield{author}{\bibinfo{person}{Steinhardt Jacob}, \bibinfo{person}{Wei~Koh
  Pang}, {and} \bibinfo{person}{Liang Percy}.} \bibinfo{year}{2017}\natexlab{}.
\newblock \showarticletitle{Certified defenses for data poisoning attacks}. In
  \bibinfo{booktitle}{\emph{Proceedings of the International Conference on
  Neural Information Processing Systems}}. \bibinfo{pages}{3520–3532}.
\newblock


\bibitem[Kolouri et~al\mbox{.}(2020)]%
        {kolouri2020universal}
\bibfield{author}{\bibinfo{person}{Soheil Kolouri}, \bibinfo{person}{Aniruddha
  Saha}, \bibinfo{person}{Hamed Pirsiavash}, {and} \bibinfo{person}{Heiko
  Hoffmann}.} \bibinfo{year}{2020}\natexlab{}.
\newblock \showarticletitle{Universal litmus patterns: Revealing backdoor
  attacks in cnns}. In \bibinfo{booktitle}{\emph{Proceedings of the IEEE/CVF
  Conference on Computer Vision and Pattern Recognition}}.
  \bibinfo{pages}{301--310}.
\newblock


\bibitem[Krizhevsky et~al\mbox{.}(2009)]%
        {krizhevsky2009learning}
\bibfield{author}{\bibinfo{person}{Alex Krizhevsky}, \bibinfo{person}{Geoffrey
  Hinton}, {et~al\mbox{.}}} \bibinfo{year}{2009}\natexlab{}.
\newblock \showarticletitle{Learning multiple layers of features from tiny
  images}.
\newblock  (\bibinfo{year}{2009}).
\newblock


\bibitem[Li et~al\mbox{.}(2021a)]%
        {li2021invisible}
\bibfield{author}{\bibinfo{person}{Yuezun Li}, \bibinfo{person}{Yiming Li},
  \bibinfo{person}{Baoyuan Wu}, \bibinfo{person}{Longkang Li},
  \bibinfo{person}{Ran He}, {and} \bibinfo{person}{Siwei Lyu}.}
  \bibinfo{year}{2021}\natexlab{a}.
\newblock \showarticletitle{Invisible backdoor attack with sample-specific
  triggers}. In \bibinfo{booktitle}{\emph{Proceedings of the IEEE/CVF
  International Conference on Computer Vision}}. \bibinfo{pages}{16463--16472}.
\newblock


\bibitem[Li et~al\mbox{.}(2021b)]%
        {li2021neural}
\bibfield{author}{\bibinfo{person}{Yige Li}, \bibinfo{person}{Xixiang Lyu},
  \bibinfo{person}{Nodens Koren}, \bibinfo{person}{Lingjuan Lyu},
  \bibinfo{person}{Bo Li}, {and} \bibinfo{person}{Xingjun Ma}.}
  \bibinfo{year}{2021}\natexlab{b}.
\newblock \showarticletitle{Neural attention distillation: Erasing backdoor
  triggers from deep neural networks}.
\newblock \bibinfo{journal}{\emph{arXiv preprint arXiv:2101.05930}}
  (\bibinfo{year}{2021}).
\newblock


\bibitem[Liu et~al\mbox{.}(2018)]%
        {liu2018fine}
\bibfield{author}{\bibinfo{person}{Kang Liu}, \bibinfo{person}{Brendan
  Dolan-Gavitt}, {and} \bibinfo{person}{Siddharth Garg}.}
  \bibinfo{year}{2018}\natexlab{}.
\newblock \showarticletitle{Fine-pruning: Defending against backdooring attacks
  on deep neural networks}. In \bibinfo{booktitle}{\emph{International
  Symposium on Research in Attacks, Intrusions, and Defenses}}.
  \bibinfo{pages}{273--294}.
\newblock


\bibitem[Moreland and Angel(2003)]%
        {moreland2003fft}
\bibfield{author}{\bibinfo{person}{Kenneth Moreland} {and}
  \bibinfo{person}{Edward Angel}.} \bibinfo{year}{2003}\natexlab{}.
\newblock \showarticletitle{The FFT on a GPU}. In
  \bibinfo{booktitle}{\emph{Proceedings of the ACM SIGGRAPH/EUROGRAPHICS
  Conference on Graphics Hardware}}. \bibinfo{pages}{112--119}.
\newblock


\bibitem[Nguyen and Tran(2021)]%
        {nguyen2021wanet}
\bibfield{author}{\bibinfo{person}{Anh Nguyen} {and} \bibinfo{person}{Anh
  Tran}.} \bibinfo{year}{2021}\natexlab{}.
\newblock \showarticletitle{WaNet: Imperceptible Warping-based Backdoor
  Attack}.
\newblock \bibinfo{journal}{\emph{arXiv preprint arXiv:2102.10369}}
  (\bibinfo{year}{2021}).
\newblock


\bibitem[Peng et~al\mbox{.}(2021)]%
        {peng2021conformer}
\bibfield{author}{\bibinfo{person}{Zhiliang Peng}, \bibinfo{person}{Wei Huang},
  \bibinfo{person}{Shanzhi Gu}, \bibinfo{person}{Lingxi Xie},
  \bibinfo{person}{Yaowei Wang}, \bibinfo{person}{Jianbin Jiao}, {and}
  \bibinfo{person}{Qixiang Ye}.} \bibinfo{year}{2021}\natexlab{}.
\newblock \showarticletitle{Conformer: Local features coupling global
  representations for visual recognition}. In
  \bibinfo{booktitle}{\emph{Proceedings of the IEEE/CVF International
  Conference on Computer Vision}}. \bibinfo{pages}{367--376}.
\newblock


\bibitem[Saha et~al\mbox{.}(2020)]%
        {saha2020hidden}
\bibfield{author}{\bibinfo{person}{Aniruddha Saha},
  \bibinfo{person}{Akshayvarun Subramanya}, {and} \bibinfo{person}{Hamed
  Pirsiavash}.} \bibinfo{year}{2020}\natexlab{}.
\newblock \showarticletitle{Hidden trigger backdoor attacks}. In
  \bibinfo{booktitle}{\emph{Proceedings of the AAAI Conference on Artificial
  Intelligence}}. \bibinfo{pages}{11957--11965}.
\newblock


\bibitem[Selvaraju et~al\mbox{.}(2017)]%
        {selvaraju2017grad}
\bibfield{author}{\bibinfo{person}{Ramprasaath~R Selvaraju},
  \bibinfo{person}{Michael Cogswell}, \bibinfo{person}{Abhishek Das},
  \bibinfo{person}{Ramakrishna Vedantam}, \bibinfo{person}{Devi Parikh}, {and}
  \bibinfo{person}{Dhruv Batra}.} \bibinfo{year}{2017}\natexlab{}.
\newblock \showarticletitle{Grad-cam: Visual explanations from deep networks
  via gradient-based localization}. In \bibinfo{booktitle}{\emph{Proceedings of
  the IEEE/CVF International Conference on Computer Vision}}.
  \bibinfo{pages}{618--626}.
\newblock


\bibitem[Stallkamp et~al\mbox{.}(2012)]%
        {stallkamp2012man}
\bibfield{author}{\bibinfo{person}{Johannes Stallkamp}, \bibinfo{person}{Marc
  Schlipsing}, \bibinfo{person}{Jan Salmen}, {and} \bibinfo{person}{Christian
  Igel}.} \bibinfo{year}{2012}\natexlab{}.
\newblock \showarticletitle{Man vs. computer: Benchmarking machine learning
  algorithms for traffic sign recognition}.
\newblock \bibinfo{journal}{\emph{Neural Networks}}  \bibinfo{volume}{32}
  (\bibinfo{year}{2012}), \bibinfo{pages}{323--332}.
\newblock


\bibitem[Tanchenko(2014)]%
        {tanchenko2014visual}
\bibfield{author}{\bibinfo{person}{Alexander Tanchenko}.}
  \bibinfo{year}{2014}\natexlab{}.
\newblock \showarticletitle{Visual-PSNR measure of image quality}.
\newblock \bibinfo{journal}{\emph{Journal of Visual Communication and Image
  Representation}} (\bibinfo{year}{2014}), \bibinfo{pages}{874--878}.
\newblock


\bibitem[Wang et~al\mbox{.}(2019)]%
        {wang2019neural}
\bibfield{author}{\bibinfo{person}{Bolun Wang}, \bibinfo{person}{Yuanshun Yao},
  \bibinfo{person}{Shawn Shan}, \bibinfo{person}{Huiying Li},
  \bibinfo{person}{Bimal Viswanath}, \bibinfo{person}{Haitao Zheng}, {and}
  \bibinfo{person}{Ben~Y Zhao}.} \bibinfo{year}{2019}\natexlab{}.
\newblock \showarticletitle{Neural cleanse: Identifying and mitigating backdoor
  attacks in neural networks}. In \bibinfo{booktitle}{\emph{Proceedings of the
  IEEE Symposium on Security and Privacy}}. \bibinfo{pages}{707--723}.
\newblock


\bibitem[Wang et~al\mbox{.}(2022)]%
        {wang2022invisible}
\bibfield{author}{\bibinfo{person}{Tong Wang}, \bibinfo{person}{Yuan Yao},
  \bibinfo{person}{Feng Xu}, \bibinfo{person}{Shengwei An},
  \bibinfo{person}{Hanghang Tong}, {and} \bibinfo{person}{Ting Wang}.}
  \bibinfo{year}{2022}\natexlab{}.
\newblock \showarticletitle{An Invisible Black-Box Backdoor Attack Through
  Frequency Domain}. In \bibinfo{booktitle}{\emph{Proceedings of the European
  Conference on Computer Vision}}. \bibinfo{pages}{396--413}.
\newblock


\bibitem[Xu et~al\mbox{.}(2020)]%
        {xu2020defending}
\bibfield{author}{\bibinfo{person}{Kaidi Xu}, \bibinfo{person}{Sijia Liu},
  \bibinfo{person}{Pin-Yu Chen}, \bibinfo{person}{Pu Zhao}, {and}
  \bibinfo{person}{Xue Lin}.} \bibinfo{year}{2020}\natexlab{}.
\newblock \showarticletitle{Defending against backdoor attack on deep neural
  networks}.
\newblock \bibinfo{journal}{\emph{arXiv preprint arXiv:2002.12162}}
  (\bibinfo{year}{2020}).
\newblock


\bibitem[Yan et~al\mbox{.}(2019)]%
        {yan2019fourier}
\bibfield{author}{\bibinfo{person}{Tao Yan}, \bibinfo{person}{Jiamin Wu},
  \bibinfo{person}{Tiankuang Zhou}, \bibinfo{person}{Hao Xie},
  \bibinfo{person}{Feng Xu}, \bibinfo{person}{Jingtao Fan}, \bibinfo{person}{Lu
  Fang}, \bibinfo{person}{Xing Lin}, {and} \bibinfo{person}{Qionghai Dai}.}
  \bibinfo{year}{2019}\natexlab{}.
\newblock \showarticletitle{Fourier-space diffractive deep neural network}.
\newblock \bibinfo{journal}{\emph{Physical Review Letters}}
  (\bibinfo{year}{2019}), \bibinfo{pages}{023901}.
\newblock


\bibitem[Yang and Soatto(2020)]%
        {yang2020fda}
\bibfield{author}{\bibinfo{person}{Yanchao Yang} {and} \bibinfo{person}{Stefano
  Soatto}.} \bibinfo{year}{2020}\natexlab{}.
\newblock \showarticletitle{Fda: Fourier domain adaptation for semantic
  segmentation}. In \bibinfo{booktitle}{\emph{Proceedings of the IEEE/CVF
  Conference on Computer Vision and Pattern Recognition}}.
  \bibinfo{pages}{4085--4095}.
\newblock


\bibitem[Zeng et~al\mbox{.}(2021)]%
        {zeng2021rethinking}
\bibfield{author}{\bibinfo{person}{Yi Zeng}, \bibinfo{person}{Won Park},
  \bibinfo{person}{Z~Morley Mao}, {and} \bibinfo{person}{Ruoxi Jia}.}
  \bibinfo{year}{2021}\natexlab{}.
\newblock \showarticletitle{Rethinking the backdoor attacks' triggers: A
  frequency perspective}. In \bibinfo{booktitle}{\emph{Proceedings of the
  IEEE/CVF International Conference on Computer Vision}}.
  \bibinfo{pages}{16473--16481}.
\newblock


\bibitem[Zhang et~al\mbox{.}(2018)]%
        {zhang2018unreasonable}
\bibfield{author}{\bibinfo{person}{Richard Zhang}, \bibinfo{person}{Phillip
  Isola}, \bibinfo{person}{Alexei~A Efros}, \bibinfo{person}{Eli Shechtman},
  {and} \bibinfo{person}{Oliver Wang}.} \bibinfo{year}{2018}\natexlab{}.
\newblock \showarticletitle{The Unreasonable Effectiveness of Deep Features as
  a Perceptual Metric}. In \bibinfo{booktitle}{\emph{Proceedings of the
  IEEE/CVF Conference on Computer Vision and Pattern Recognition}}.
  \bibinfo{pages}{586--595}.
\newblock


\bibitem[Zhang et~al\mbox{.}(2021)]%
        {zhang2021deep}
\bibfield{author}{\bibinfo{person}{Xingxuan Zhang}, \bibinfo{person}{Peng Cui},
  \bibinfo{person}{Renzhe Xu}, \bibinfo{person}{Linjun Zhou},
  \bibinfo{person}{Yue He}, {and} \bibinfo{person}{Zheyan Shen}.}
  \bibinfo{year}{2021}\natexlab{}.
\newblock \showarticletitle{Deep stable learning for out-of-distribution
  generalization}. In \bibinfo{booktitle}{\emph{Proceedings of the IEEE/CVF
  Conference on Computer Vision and Pattern Recognition}}.
  \bibinfo{pages}{5372--5382}.
\newblock


\bibitem[Zhong et~al\mbox{.}(2022)]%
        {DBLP:conf/ijcai/ZhongQZ22}
\bibfield{author}{\bibinfo{person}{Nan Zhong}, \bibinfo{person}{Zhenxing Qian},
  {and} \bibinfo{person}{Xinpeng Zhang}.} \bibinfo{year}{2022}\natexlab{}.
\newblock \showarticletitle{Imperceptible Backdoor Attack: From Input Space to
  Feature Representation}. In \bibinfo{booktitle}{\emph{Proceedings of the
  International Joint Conference on Artificial Intelligence}}.
  \bibinfo{pages}{1736--1742}.
\newblock


\end{thebibliography}










\end{document}